\input harvmac

\vskip 1cm

 \Title{ \vbox{\baselineskip12pt\hbox{  Brown Het-1159 }}}
 {\vbox{
\centerline{ Non-commutative gravity from }
\centerline{ the  ADS/CFT correspondence.  }  }}

\centerline{$\quad$ { Antal Jevicki and Sanjaye Ramgoolam }}
\smallskip
\centerline{{\sl  }}
\centerline{{\sl Brown  University}}
\centerline{{\sl Providence, RI 02912 }}
\centerline{{\tt antal,ramgosk@het.brown.edu}}
 \vskip .3in 
 
 The exclusion principle of Maldacena and Strominger
 is seen to 
 follow from deformed Heisenberg algebras associated with 
 the chiral  rings of  $S_N$ orbifold CFTs.
 These deformed algebras are related to 
 quantum groups at roots of unity, and 
 are interpreted as  algebras of  space-time 
 field creation and annihilation operators. 
 We also propose, as space-time 
 origin of the stringy exclusion principle, 
 that the $ADS_3 \times S^3$ 
 space-time of the associated six-dimensional supergravity theory 
 acquires, when quantum effects are taken into account,  
 a non-commutative structure given by 
 $ SU_q(1,1) \times SU_q (2)$.  Both  remarks imply that 
 finite $N$ effects are captured by quantum groups $SL_q(2)$ 
 with $q=  e^{ { i  \pi \over { N + 1 } } }$.
 This implies that a proper  framework for 
 the theories in question is given by gravity  
 on a non-commutative  spacetime with 
 a  q-deformation  of field oscillators.
 An interesting consequence of this 
 framework is a holographic 
 interpretation for a product structure
 in the space of all unitary representations of the non-compact quantum 
 group $SU_q(1,1)$ at roots of unity.


\Date{11/98}

\def\nh{ {\hat n }}
\def\aq{ {a_{q}} }
\def\aqd{ {  a_{q}^{\dagger} }}
\def\ad{  { a^{\dagger} } }  
\def\apd{  { a_{p}^{\dagger} }  }

\def\snm{ \sigma_{i_1 \cdots i_{N-1} } }

\lref\mmp{ S. Meljanac, M. Milekovic, S. Pallua, 
           { ``Unified view of deformed single oscillator algebras } 
           Phys. Lett. B 328 (1994) 55-59 }
\lref\qoshopf{ Hong Yan, `` q-deformed oscillator 
            algebra as a quantum group,''   
         J. Phys. A : Math. Gen. 23 (1990) L1155-L1160 }
\lref\gubs{ S. Gubser, {``Can the effective string see higher partial 
                         waves ?''} Phys. Rev. D56: 4984-4993, hepth/9704195 }
\lref\dsww{ Dine,  Seiberg, Wen. Witten } 
\lref\vq{ Vafa, ``Quantum Rings,''}  
\lref\witcs{ Witten on $k+2$ } 
\lref\Pit{ L. Pitner, ``Algebraic foundations of Non-commutative 
    differential geometry and quantum groups.'' Springer-Verlag, 1996.  } 
\lref\malda{ J. Maldacena, `` The large N limit of superconformal 
                             field theories and supergravity,'' 
                      Adv.Theor.Math.Phys.2: 231-252, 1998, 
                       hepth/9711200 }
 
\lref\connes{ A. Connes, ``Non-commutative Geometry,''
             Academic Press, 1994. } 
\lref\cds{ A. Connes, M. Douglas and A. Schwarz, ``Non-commutative 
            Geometry and Matrix Theory : Compactification on tori.''
           JHEP 02 (1998) 003.  } 
\lref\madf{ J. Madore, ``Gravity on fuzzy spacetime'' gr-qc/9709002 }

\lref\macf{ A. J. Macfarlane, `` On Q analogs of the quantum harmonic 
                 oscillator  and the quantum group $SU(2)_Q$''
         J. Phys. A22: 4581, 1989  } 
\lref\aa{ Antal Jevicki and Andre van Tonder, ``Finite [ q-oscillator ] 
                                   Description of 2D string theory,''
Mod.Phys.Lett.A11:1397-1410,(1996),hep-th/9601058. } 
\lref\wor{ S.L. Woronowicz, ``Unbounded elements Affiliated 
                              with $C^*$ algebras and Non-Compact 
                              quantum groups,'' CMP 136, 399 (1991) } 
\lref\suz{ T.  Matsuzaki and T. Suzuki, 
`` Unitary highest weight representation of $U_qSU(1,1)$ 
                 when $q$ is a root of unity,''
                J. Phys. A : Math. Gen. 26 (1993) 4355-4369 }   
\lref\lvw{ W. Lerche, C. Vafa, N. Warner, 
{``Chiral rings in $N=2$ superconformal theories,''}
 Nucl. Phys. B324: 427, 1989. }  
\lref\worcomp{ 
S.L. Woronowicz , `` Twisted $SU(2)$ Group. 
An example of a non-commutative 
         differential 
calculus,'' Publ. RIMS, Kyoto Univ. 23 (1987), 117-181 }  
\lref\wit{ E. Witten, 
{`` The ADS/CFT 
correspondence and topological field theory''} , hep-th/9812012} 
\lref\malfis{ D. Berenstein, R. Corraddo, J. Maldacena, W. Fischler, 
{`` The operator product expansion for Wilson loops and surfaces
 in the large N limit,''} hepth/9809188 }   
\lref\ohsi{ C. Oh,  K. Singh,  
``Real forms of the Oscillator
 quantum algebra 
 and its representations, '' Let. in Math. Phys. 36, 77-85, 1995 } 
\lref\fink{ R. Finkelstein, ``q-gauge theory,'' q-alg/9506014. } 
\lref\liyon{ M. Li and T. Yoneya, 
{``Short-Distance
 Space-Time Structure and Black Holes in String Theory : A Short
     Review of the Present Status,''} hep-th/9806240,  
  to appear in the special issue of the Journal of Chaos,
     Solitons and Fractals on "Superstrings, M, F, S...Theory"} 
\lref\dj{ A. Das and A. Jevicki, 
`` String Field theory and Physical interpretation 
             of ``D=1'' strings ''  Mod. Phys. Lett. A5; 1639-1650, 1990 } 
\lref\withol{ E. Witten, 
{``Anti-de-Sitter space and holography,''} hepth-9802150 } 
\lref\sch{ K. Schmudgen, ``Integrable Operator representations of $R^2_q$, 
                   $X_{q,\gamma}$ and $SL_q(2,R)$,'' CMP. 159 (1994) 217 } 
\lref\matsue{T.  Matsuda , K. Mimachi, 
    Y. Nagakami, M. Noumi, Y. Saburi, K. Ueno,
              {``Unitary Representations of the quantum group
               $SU_q(1,1)$: Structure of the Dual Space of $U_qSl(2)$ }
                  Lett. Math, Phys Vol. 19, 187 and  195   }

\lref\gkp {S.S.Gubser,I.R.Klebanov and A.M.Polyakov, 
{``Gauge Theory Corellators from Non-Critical String Theory,''}
hepth-9802109 }
\lref\suzi{ T. Suzuki, {``A note on the Quantum Liouville theory 
              via quantum group: an approach to strong coupling 
              Liouville theory,''} Nucl. Phys. { \bf B492}  (1997) 717-742. }  
\lref\dfms{ L. Dixon, D. Friedan, E. Martinec and S. Shenker
            ``The conformal field theory of orbifolds,'' 
            Nucl. Phys. B282 (1987) 13. }


\lref\babel {O. Babelon, { Phys. Lett.} {\bf B215} (1988) 523;\hfill\break
J-L Gervais, { Commun. Math. Phys.} {\bf 130} (1990) 257.\hfill\break
L.Alvarez-Gaume,C.Gomez and G.Sierra,{ Nucl.Phys.} {bf B319} (1989)155
\hfill\break
G.Moore and N.Reshetikin,{ Nucl.Phys.} {bf\ B328}(1989) 557 }

\lref\ldf {A. Alekseev and S. Shatashvili, { Commun. Math. Phys.} {\bf 133} (1990) 353;\hfill\break
A. Alekseev, L. Faddeev and M. Semenov-Tian-Shansky, 
{ Commun. Math. Phys.} {\bf 149} (1992) 335.\hfill\break
K. Gawedzki, { Commun. Math. Phys.} {\bf 139} (1991) 201;\hfill\break
M. Chu and P. Goddard, { Nucl. Phys.} {\bf B445} (1995) 145.\hfill\break
 P. Furlan, L. K. Hadjiivanov and I. T. Todorov,
 {``Canonical Approach to the Quantum WZNW Model,"} 
ICPT Trieste preprint IC/95/74, ESI 234 (1995).}
\lref\gopvaf{ R. Gopakumar and C. Vafa, {``On the gauge theory geometry 
 correspondence,''} hep-th/9811131 }
\lref\perr{ } 
\lref\har{ Harold Steinacker, 
{ ``Finite dimensional unitary representations 
  of quantum anti-de Sitter groups,''}
   { Commun. Math. Phys.}  { \bf 192} 687-706, 1998 } 
\lref\cmr{S. Cordes, G. Moore, S. Ramgoolam, { ``Large N 2D Yang Mills
       Theory and Topological String Theory,''} 
       hep-th/9402107, CMP 185, 543, 1997} 
\lref\grta{ D. Gross and W. Taylor, { `` Two Dimensional QCD 
         is a string theory,''} 
         { Nucl. Phys.}  {\bf B400 } (1993) 181, hep-th/9301068 } 
\lref\dm{ R. Dijkgraaf and G. Moore, 
        { ``Balanced topological field theories,''} 
        CMP 185, 411-440, 1997, hep-th/9608169. } 
\lref\hortwod{ P. Horava, { `` Topological rigid string theory 
                     and two-dimensional QCD,''} hep-th/9507060,
                      { Nucl.Phys.} {\bf  B463 } (1996) 238 }
\lref\podles{ P. Podles, { `` Quantum spheres,''} 
      { Lett. Math. Phys.}  14 : 193-202, 1987 } 
\lref\dijk{ R. Dijkgraaf, { ``Instanton strings and Hyperkahler
    geometry,''} hep-th/9810210.  }
\lref\dvvm{ R. Dijkgraaf, G. Moore, E. Verlinde, H. Verlinde,  
    {`` Elliptic genera of Symmetric products and Second quantized strings,''} 
    hep-th/9608096, Commun. Math. Phys. 185, (1997) 197-209.  } 
\lref\witcfth{E. Witten, { `` On the conformal field theory 
             of the Higgs branch,'' } hep-th/9707093, JHEP. 07 (1997) 003 }

\newsec{ INTRODUCTION } 

 The ADS/CFT \malda\gkp\withol\ correspondence  
 gives a large class of
 examples where the existence 
 of a  well defined theory of quantum gravity 
 follows from the existence of  a dual
 conformal field theory. Often we have access
 to regimes where perturbative string theory 
 is not valid.   The usual 
 stringy explanations for the finiteness 
 of the quantum gravity, like an infinite tower of 
 massive states, are therefore  not directly applicable. 
 We can ask what, from the 
 spacetime point of view, are the mechanisms that 
 lead to a finite quantum theory of gravity.

 \nref\malstrom{ J. Maldacena, A. Strominger, ``ADS3 black holes 
                   and a Stringy Exclusion Principle,'' hepth/9804085.   } 

 \nref\suswi {L.Susskind and E.Witten, 
 {``The Holographic Bound of Anti-de Sitter
 Space,``}hep-th/9805114,  \hfill\break
 A.W.Peet and J. Polchinski, { `` UV/IR relations 
 in ADS dynamics,'' hep-th/9809022.  }   }  

 \nref\jevyon{A. Jevicki and T. Yoneya, ``Spacetime Uncertainty principle and
                                     Conformal Symmetry in D-particle 
                    dynamics,''   Nucl.Phys. B535 (1998) 335-348, 
hep-th/9805069 ;\hfill\break
A.Jevicki,Y.Kazama and T.Yoneya,{``Generalized Conformal Symmetry in D-Brane
Matrix Models``,}hep-th/9810146.}

\nref\deger{S. Deger, A. Kaya, E. Sezgin and P. Sundell, 
{``Spectrum of $D=6, N=4B$ Supergravity 
on Ads in Three-Dimensional $ADS_3 \times S^3$,"} hep-th/9804166.}
\nref\deboer {J. de Boer, {``Six-Dimensional 
Supergravity and 2-D Conformal Field,"} hep-th/9806104.}
\nref\dbii{ J. de Boer, {`` Large N elliptic genus and 
                ADS/CFT correspondence,''}  hep-th/9812240.  } 
\nref\many{C. Vafa, {``Puzzles at Large N,"}hep-th/9804172, \hfill\break
 E.Martinec, {``Matrix Models of AdS Gravity``} hepth/9804111, \hfill\break
 V. Balasubramanian, P. Kraus and A. Lawrence, 
   {``Bulk vs. Boundary Dynamics in Anti-De-Sitter Space-Time,"} 
    hep-th/9805171 \hfill\break
F.Larsen, {``The Perturbation Spectrum of Black Holes
in N=8 Supergravity,``} hep-th/9805208, \hfill\break
J. M. Evans, M. R. Gaberdiel and M. J. Perry,
 {``The No Ghost Theorem for ADS(3) and the Stringy Exclusion Principle,"}
 hep-th/9806024 \hfill\break
A. Giveon, D. Kutasov and N. Seiberg, 
{``Comments on String Theory on ADS(3),"} hep-th/9806194, \hfill\break
J. R. David, G. Mandal and S. R. Wadia, 
 {``Absorption and Hawking Radiation of Minimal and Fixed Scalars,
 and ADS/CFT Correspondence,"} hep-th/9808168\hfill\break
D.Kutasov,F.Larsen,R.Leigh,{``String Theory in Magnetic
Monopole Backgrounds``} hep-th/9812027\hfill\break
J.de Boer,H.Ooguri,H.Robins and J.Tannenhauser, {``String
 Theory on $AdS_3$,``} hep-th/9812046. } 


  Numerous recent studies have given considerable insight
 into the nature of the ADS/CFT correspondence. Some of the novel           
 features of the emerging spacetime theory  can be 
 seen in  \refs{ \malstrom - \jevyon }. 
 For investigation of many of the outstanding questions 
 regarding 5D black holes related to a system 
 of $Q_1$ D1 and  $Q_5$ D5-branes 
 the CFT of $S^N (X)$ ( $X$ is $K3$ or $T^4$ ),
 where $ N= Q_1Q_5 $,  provides a particularly
 useful  laboratory\refs{\deger,\deboer, \dbii, \many }.
 These orbifold CFTs have played  important and diverse
  roles in recent works, see for example 
 \refs{\dvvm, \dijk, \witcfth }.
 
 One of the most interesting phenomena found
 in these studies is 
 a  stringy exclusion principle  
 \malstrom, related to  the unitarity 
  of representations of the superconformal algebra in the 2D 
   CFT with target space $S^N( X )$. 
  The $N$ parameter here is analogous to 
  the $N$ of $SU(N)$ Yang Mills theory 
 in the $ADS_5 \times S^5$ case. Here $1/N$ 
 is the expansion parameter for the semiclassical gravitational 
expansion for 6D gravity on $ADS_3 \times S^3$.    
 The question that we address in the present work
 concerns  the spacetime explanation of this exclusion principle
 and its meaning
 at the level of supergravity.
  We argue that the classical 
  $ADS_3 \times S^3$ spacetime should be understood as a non-commutative 
  manifold $ SU_q(1,1) \times SU_q(2)$, with $q$ a root of unity, 
  when quantum effects are taken into account. 
 We also suggest that an identical deformation applies to
 the (supergravity) field oscillators themselves. 
 Earlier, an equivalent role of
 q-oscillators  at roots of unity was proposed  in \aa, 
 in the context of two dimensional string theory\dj.  
 We are now lead  to suggest that this represents a general feature of the 
 matrix-spacetime correspondence.

 In general a systematic way to define deformations 
 of spacetime is studied  under the heading 
 of non-commutative geometry \connes. 
 Gauge theory on non-commutative  
 spaces has found a natural role in D-brane physics
 recently \cds.
 There is also a large literature attempting 
 to define gravity in the non-commutative setting  
 (for a review of the successes and difficulties 
 see for example \madf). The version of gravity that we
 propose from CFT  has some elements of similarity
 with the above literature, but also some differences. One
 characteristic feature implied in the present (and also
 other) non-commutative 
 representations is the appearance of a
 space-time uncertainty relation \liyon.
 
 The plan of the paper is the following.  
In section 2, we derive
 several important algebraic properties 
of the chiral ring of $S^N(X)$. For concreteness
 we work with $X= T^4$ but the arguments are general. 
   We describe in a
 simple setting how the implementation of
 the exclusion principle 
 requires that the creation and annihilation 
 operators of  
 fields obey a modified Heisenberg algebra. The relevant 
 Heisenberg algebras are found from  
 the CFT dual. They are shown to have a representation
 in terms of q-oscillators \macf\  with 
 $q =  e^{ { i  \pi \over { N + 1 } } }$. 
 The $q$-oscillators are also related to $SL_q(2)$ ( Note 
 that in discussions where the real form 
 does not matter much we use this notation, 
 otherwise we will distinguish between
 $SU_q(2)$ and $SU_q(1,1)$ and even $SL_q(2,R)$). 
 At large $N$, we have the ordinary 
 Heisenberg algebra.
 The discussion in the first part 
 of section 2 is 
 restricted to a simple 
 class of field operators which, in the CFT, 
 are related to the untwisted sector. 
 We proceed to 
 give some details on the construction 
 of  twisted sector operators. The  
 same qualitative properties and 
 deformation of Heisenberg algebras 
 also survive when we consider the 
 operators from the twisted sector. 

The final part of section 2 
derives the detailed form of the exclusion principle
on the generators of the chiral primaries.  
 Although the general fact of the existence 
 of the exclusion principle follows from the unitarity 
 of the superconformal algebra, the detailed form of the 
 exclusion principle coming from the orbifold CFT 
 is not predicted by this argument.
 Some interesting 
 new features are derived and turn out to 
 be precisely captured by the idea of a quantum group 
 symmetry, as elaborated in section 3. 
 The results can be expressed simply in terms  
 of a bound of the left and right $SU(2)$ quantum numbers
  of   the chiral primaries.

 Section  3 deals with a spacetime interpetation of these
 results.  It was shown in \deboer\ that 
 KK reduction of supergravity fields 
 matches the spectrum of chiral primaries at large $N$. 
 An ordinary KK reduction 
 on $S^3$ produces an infinite number of 
 representations of $SU(2)$. The truncation 
 of the number of unitary  representations 
 is familiar in going from the 
 universal enveloping algebra  $U(SU(2))$
 to $U_qSU(2)$. The deformation of the algebra 
 can be related to a deformation of the $SU(2)$ manifold
 to  a non-commutative manifold $SU_q(2)$ \worcomp. 
 In terms of the geometry of $SU_q(2)$ the 
 cutoff on the spectrum of unitary reps 
 of $U_qSU(2)$ means that Fourier transformation 
 on the deformed manifold involves a finite set 
 of $SU(2)$ representations. (  In the rest of this paper, we will not  
 always be  careful to use separate notation 
 for the q-deformed algebra and the dual q-deformed
 manifold, using $SU_q(2)$ for both, since it will often 
 be clear from the context which one is being refered to. )
 The detailed form of the CFT cutoffs, derived 
in section 2,  on the chiral primaries
 takes the precise form expected from KK reduction on 
 $SU_q(2)$ with $q^{ \pi i \over { N +1  } } $, rather than 
 $SU(2)$.

 The next part of section 3
 shows that not only  
 $SU(2)= S^3$ is deformed to $SU_q(2)$ but that 
 the $ AdS_3 = SU(1,1)$ is also deformed  
 to $SU_q (1,1)$, again with the same value of $q$. 
 This allows a discussion of the cutoffs on the
 superalgebra  descendants
 of chiral primaries. 
 Since we are claiming that 
 the CFT living on an ordinary commutative Riemann 
 surface is equivalent to gravity on
 a non-commutative spacetime, consistency 
 with holography requires  ordinary Riemann 
 surface to appear in the boundary of $SU_q(1,1)$. 
 A result of \suz\ allows us to prove that it does. 
 This discussion of holography requires, 
 not surprisingly, a consideration of operators
 which allow the creation of states of sufficiently
 large mass that they can change the background into one 
 with black holes. 
 
 We observe that the $q$ values obtained
 from the deformed Heisenberg algebra 
 and from the above $q$-deformation of $SU(2)$ are identical. 
 The agreement of the $q$-parameters from two different 
 kinds of physics leads to a speculation 
 on algebraic structures associated to the chiral ring
 and their space-time interpretation.    
 We then  outline
 an approach,  based on the ordinary WZW CFT/quantum group 
 correspondence,  to make the presence of non-commutative 
 spacetime coordinates explicit. In the final part of
 section 3, we attempt  
 to get a glimpse of the possibilities 
 coming from generalizing the idea of $q$-deformed 
 spacetime to other ADS backgrounds entering 
 the ADS-CFT correspondence.

 We conclude with a summary and
 outline possible strategies to improve the 
 understanding of the relevance of gravity 
 in the non-commutative setting to the 
 elucidation of quantum effects in the ADS-CFT 
 correspondence.

\newsec{ Algebraic properties of chiral ring of SCFT on $S^N(T^4)$ } 

The spacetime fields in $ADS_3 \times S^3$ 
supergravity 
are related to chiral primary operators 
in the dual CFT. A detailed discussion
of the spacetime physics requires 
a derivation of several properties of 
the chiral ring, directly from the CFT. 
The properties of interest fall into two broad 
categories. One concerns the nilpotence 
of the generators of the chiral ring 
and the second concerns the truncation in the 
number of generators. Sections 2.1 and 2.3
 concern the former property. 
Section 2.4 concerns the latter.
Section 2.2 contains some technical points
necessary for the subsequent discussion. 

\subsec{ Deformed oscillator algebra and $SL_q(2)$ }  

 In the superconformal field with the symmetric product target 
 space $S^N(T^4)$ , consider the $S_N$ invariant operator 
 related to a $(1,1)$  form on $T^4$ of the form $\psi \bar \psi$. 
\eqn\op{\eqalign{ &  \alpha_{-1} = \sum_{i=1}^{N} \psi_{i} \bar \psi_{i}  \cr 
                  &  \alpha_{1} = \sum_{i=1}^{N} { \bar \psi_{i} }^{\dagger }
                 {  \psi_{i}}^{\dagger } \cr } }
 The two fermions are each chosen from one of two complex fermions, 
 from the left moving and the right moving part of the CFT 
 respectively. 
The anti-commutators are 
\eqn\coms{\eqalign{ & \{ \psi_i , \psi^{\dagger }_j  \} = \delta_{ij} \cr 
                    & \{ \bar \psi_i , { \bar \psi}_j^{\dagger } \} =
                     \delta_{ij} \cr }} 
To make the connection with CFT we note the following 
\eqn\cftmap{\eqalign{ 
          & \psi = \psi_{-1/2} \cr 
          & \psi^{\dagger } = \psi^{*}_{ 1/2} \cr 
          & \bar \psi = \bar \psi_{-1/2} \cr  
           & {\bar \psi}^{\dagger} =  \bar \psi^{*}_{1/2} \cr }}
In the CFT, the fields $\psi, \psi^*$ are left movers
( depending on $z$ ) while $\bar \psi, \bar \psi^*$ 
are right movers (depending on $\bar z$ ).
Note that $\psi \bar \psi$ belongs to the 
$(c,c)$ ring. $\psi$ is annihilated by the $SU(2) $ 
raising operator, $\psi^{\dagger} $ by the lowering operator. 
The product is not annihilated by either. 
        
The commutator of the bosonic oscillators gives:
\eqn\bos{ 
[  {1 \over {\sqrt N}}\alpha_{1}, {1 \over {\sqrt N} }\alpha_{-1} ] = 
  1 - {1\over N } \sum_{i} ( \psi_{i} \psi^{\dagger}_{i} + 
      \bar \psi_{i}  {\bar \psi}^{\dagger}_{i} )  }
In the Hilbert space obtained by acting with 
the creation operators $\psi$ on a vacuum annihilated
by the $\psi^{\dagger}$ find that the second term acts 
as a number operator. Indeed 
defining 
$$ \hat n = { 1\over 2 }  \sum_{i}  ( \psi_{i}\psi^{\dagger}_{i} + 
      \bar \psi_{i} {\bar \psi}^{\dagger}_{i}   ) . $$
\eqn\num{
[  \hat n , \alpha_{\pm 1 }  ] = \pm  \alpha_{\pm 1 },   } 
we can rewrite \bos\ as
\eqn\bosi{ [  a_p , \bar a_p ] = 1 - {2 {\hat n} \over N} } 
This algebra has been studied as a deformation 
of the Heisenberg algebra and has been named the 
parafermion algebra ( see \mmp\ and earlier refs there ). 
One has the relation 
\eqn\rel{  \phi_p (\nh ) = \bar a_p a_p = \nh - {\nh(\nh -1)\over N}.  } 
and also that $\phi_p (\nh +1) = a_p \bar a_p$. 
We have shown that the chiral  ring ( and its $SU(2)$ descendants )   
has relations of the same form as the parafermion algebra.

 Another oscillator algebra which is related to the 
 ring of chiral operators and their conjugates is the 
 q-oscillator algebra, for $q$ a root of unity.
  The relevance of the q-deformed oscillator algebra
 is easily guessed. Because the $\alpha_{-1}$ is constructed
 in terms of fermions, it satisfies $\alpha_{-1}^{N +1} = 0 $. 
  The usual Heisenberg algebra relation 
\eqn\us{ [ a , \ad ] = 1 } 
is not compatible with this nilpotence.
 A deformation which is compatible is the 
 q-oscillator algebra 
\eqn\qosc{ \aq \aqd - q^{-1} \aqd \aq = q^{ \hat n },  } 
with $ q = e^{ i \pi \over { N + 1 }} $.  
 Indeed it can be checked that 
 \eqn\chk{ \aq (\aqd )^{N+1} = - (\aqd )^{N+1} \aq. }
 So imposing the nilpotence is consistent with the 
 relations of the algebra. 
 The q-oscillator algebra can be related to 
 $SL_q(2)$ \macf\ by a map in which $\aq^2$ is related 
 to $X_{-}$, ${\aqd}^2 $ is related to 
 $X_{+}$ and $\hat n$ is related to $H$.

 A precise connection to the q-oscillator algebra, 
 which we will develop, allows us to associate   
 a Hopf algebra structure to the 
 set of operators generated by the chiral operators and their conjugates. 
 This will use results on the Hopf algebra structure 
 of q-oscillators. Ref. \qoshopf\ gives a 
 co-product $\Delta$, an antipode $S$, and an R-matrix
 for the q-oscillators. This allows us to associate all 
 these structures to the algebra of chiral operators and their conjugates.

 We can transform from the oscillators in \op\ 
 to the q-oscillators. A crucial role is played by the
 functions $\phi_p ( \hat n )$ and $\phi_q ( \hat n ) $
\eqn\phieq{\eqalign{ & 
 \phi_{p} ( \hat n ) = \nh ( 1 - { 1\over N}( \nh - 1 ) ) \cr 
  & \phi_{q} ( \nh ) = { { q^{ \nh } - q^{-\nh } } \over { q - q^{-1} } } 
 \cr } }
 To all orders in $1/N$, 
 they can be used to map the parafermion oscillators 
 and the $q$-oscillators to  
 the ordinary Heisenberg algebra \mmp. 
To map between parafermion oscillators and Heisenberg algebra, we use 
\eqn\apth{ a_{p} = a 
    \sqrt { \phi_{p} ( \nh ) \over{ N  } } }
 This map is singular at finite $N$ because 
 $\phi_{p} ( \nh = N+1 )  = 0 $.   
To map between q-oscillators and Heisenberg algebra 
 we use 
\eqn\aqth{ a_{q} = a 
    \sqrt { \phi_{q} ( \nh ) \over{ N  } } }
 This is also well-defined in the large  $N$ 
 expansion but becomes singular precisley at finite $\nh = N+1$, 
 if we choose $q = e^{i\pi \over { N+1}}$.  
 To map between parafermion algebra and $q$-oscillator algebra 
\eqn\map{ a_{p} = a_q 
          \sqrt { \phi_{p} ( \hat n ) \over{  \phi_{q} ( \hat n ) }}
} 
This map is well defined for values of $\nh $ between 
 $0$ and $N+1$.
 These are the only values of interest 
 because the $(c c^{\dagger} )$ ring is just the 
 quotient of the parafermion algebra by the 
 relation $ a_{p}^{(N+1)} = 0, \apd^{ N+1} =0 $.
 This quotient algebra can be mapped to the 
 quotient of the q-algebra by the analogous relations
  $$ a_{q}^{ (N+1 )} = 0, \quad
     \aqd^{ (N+1)} = 0. $$

Using this map to q-oscillators we can write 
 down the co-product for the algebra over the complex numbers  
 generated by the $c$ and $c^{\dagger}$.
 For example 
\eqn\cop{\eqalign{ 
&   \Delta ( a_p ) = \Delta ( a_q 
 \sqrt { \phi_{p} ( \hat n ) \over{  \phi_{q} ( \hat n ) }}  ) \cr 
&                            = \Delta (a_q) \Delta 
 (  \sqrt { \phi_{p} ( \hat n ) \over{  \phi_{q} ( \hat n )}}  ) \cr }}
 where we used the algebra homomorphism property of the 
co-product.

 To recapitulate, we have invoked q-oscillators
 as  the underlying explanation of the 
 nilpotence of the generators of the chiral ring.
  Note that this dervation of the cutoff on the 
   number operators of the generators of the chiral 
   ring has a similar flavour to the standard derivation 
    based on the chiral superconformal algebra, 
     but has important differences since the $\alpha$
      are generically  non-chiral objects from the worldsheet 
       point of view and have a spacetime interpretation in terms of 
        field oscillators. 
  In general our CFT generates a sequence of oscillators
 labeled by the harmonic forms and an index labelling cycle lengths. 
 A more careful look at the deformation of the 
 algebra shows that the deformation terms  
 appearing in the commutator of 
 $ \alpha_{1}^{1,1}$ with $\alpha_{-1}^{1,1}$ 
involve also the other operators 
which count the number of oscillators of other types. 
Requiring the positivity of 
 $< \alpha_{-1} \alpha_{1}> $ in states with various oscillator numbers 
imposes constraints on the values that these number 
operators can take at finite $N$. 
For example the terms involving the pure twist operators take the form 
given below : 
\eqn\extos{ [ \alpha_{1}^{1,1} , \alpha_{-1}^{1,1} ] 
            = 1 - { 2 \over N } ( N_1 +  \sum_{l} a_l |O_l^{\prime}> < O_l | 
             + \cdots )  }  
where the $ a_l $ are some positive coefficients 
growing with the twist number
$l$.
We will not try to 
unravel these constraints here, which will in general involve
 not just chiral primaries but also their descendants.
 We will turn to a more direct 
 determination of the constraints 
 on the operators of greatest interest, the generators
 of the chiral ring, in section (2.4).

\subsec{ Twist Operators } 
There are twist operators \dfms\ 
 for a 
boson which have conformal dimension given by  : 
 \eqn\dimflds{ \Delta ( \sigma_{k,n} (X) ) = 
{k \over {2n }} (1 - {k\over {n}} ).  } 
These satisfy the condition 
\eqn\defkn{ \partial X (z) \sigma_{k,n}( X)(0)  = 
z^{-( {k\over n} -1 ) } \tau_{k,n} + \cdots }
so that under transport around this twist operator 
we have the monodromy
\eqn\trans{ X \rightarrow e^{2i k  \pi\over { n} } X. }
 
For fermions, we have 
$\sigma_{k,n} ( \psi )$, which have dimension 
\eqn\dimflds{ \Delta ( \sigma_{k,n} (\psi ) ) = 
{k^2\over {2n^2 }}   } 
and satisfy 
\eqn\defpskn{ \psi_{(k)} (z) \sigma_{k,n}( \psi )  = z^{k/n } \tau + \cdots }  
After bosonization, 
 \eqn\twist{\eqalign{ & \psi_{(k)} = e^{i \phi_{(k)}} \cr 
    & \sigma_{k,n} (\psi_{(k)} ) = e^{i {k\over n} \phi_{(k)} }  \cr }}
We have formed linear combinations  
\eqn\lincomb{ \psi_{(k)} = 
\psi_1 + \omega^{k} \psi_2 + \cdots \omega^{kn} \psi_n }
which transform by a phase 
$\omega^{k}$ under the action of the 
n-cycle permutation. The boson $\phi_{(k)}$ bosonizes 
the pair $\psi_{(k)}$ and $\bar \psi_{(k)}$.  
Take the operator 
\eqn\fermtwist{
\sigma_n ( \psi ) = \prod_{k=1}^{n-1} e^{ i {k \over  n } \phi_{(k)} } } 
It has conformal weight $ {n \over 6} + {1 \over 12 n} - {1\over 4} $.

 For the torus $T^4$ we have 
two complex bosons and two complex fermions
so the operators 
$\alpha_{-n}( X^i, \psi^i ) $ are  built 
by combining twist operators for the bosons 
and fermions, with the index $i$ running from $1$ to $2$ :  
\eqn\osc{ \alpha_{-n} ( X^i, \psi^{i} ) = 
\prod_{i,k} e^{i { k\over n }  \phi^{i}_{(k)} } 
\prod_{i,k} \sigma_{k,n} (X^i_{(k)} )}
For two complex fermions, the conformal weight 
is $ {n \over 3 } + {1 \over 6 n} - {1\over 2}$. 
For the bosons we have $ { n \over 6 }  - { 1 \over 6 n}$. 
 Adding up the dimensions of the bosonic
 and fermionic twist operators, 
we get the weight $(n-1)/2$. 
The operator 
$e^{i { k\over n}  \phi_{(k)}} $ has a $U(1)$ 
charge of $ k/n$. After summing over $k$ we get 
a $U(1)$ charge of $(n-1)$. 
Using the expressions for the left 
moving $SU(2)$ currents ( and the analogous ones for the right-moving 
 ones ) we can easily check that the above operator 
 is annihilated by $J_0^+$ : 
\eqn\curs{\eqalign{ 
& J^{0} = \psi^{1} ( \psi^{1})^* + \psi^2 ( \psi^2)^* \cr 
& J^{+} = \psi^{1} \psi^{2} \cr 
& J^{-} = (\psi^{1})^* ( \psi^{2})^* \cr }}  
This confirms we have a chiral primary  obeying 
 as it should,  $L_0 = 1/2 J_0$ \lvw.    

We have constructed above the twist operator 
for a $Z_n$ theory. In the  case of $S_N$ we 
just sum the $Z_n$ twist operator
over all combinations 
of $n$ variables
\eqn\sumovN{ \alpha_{-n}^{(0,0)}( z,\bar z ) = \sum_{i_1 \cdots i_n} 
\sigma(  X^{i}_{i_1 \cdots i_n}, \psi^{i}_{ i_1 \cdots i_n }  ) }
 The indices $i_1 \cdots i_n$ are  $n$ distinct 
 numbers between 
 $1$ to $N$.    
 The most general chiral primary can be written in terms
 of products of operators $\alpha_{-n}^{(p,q)}$ associated
 with a $(p,q)$ form $\omega^{(p,q)} (X^i, \psi^i )$ which is 
\eqn\gencop{ \alpha_{-n}^{(p,q)} = 
  \sum_{i_1 \cdots i_n} 
\omega^{(p,q)}( X^i_{i_1}+ \cdots X^i_{i_n}, 
        \psi^i_{i_1} + \cdots \psi^i_{i_n}) 
    \sigma(  X^{i}_{i_1 \cdots i_n}, \psi^{i}_{ i_1 \cdots i_n }  ) } 
 
Schematically we will write, using a factorization of the
 $Z_n$ twist operators in terms of chiral and anti-chiral 
 operators,  
 \eqn\moreq{ \alpha_{-n}( z, \bar z ) =
 \sum_{I_{n}} O^{I_n} ( z ) \bar O^{I_n} (\bar z )   }
where $O^{I_n}$ depends on $X$ and $\psi$ fields labeled 
 by $I_n$,  which 
equals a set of indices  $(i_1, \cdots i_n)$.  
The  indices 
$i_1, i_2 \cdots i_n$,  run over $n$ { \it distinct}  numbers 
between $1$ to $N$. 
This acts on the vacuum to give a state 
\eqn\alph{\eqalign{&  
 Lim_{ z, \bar z \rightarrow 0 }   \alpha_{-n} (z , \bar z ) |0> \cr 
    &        = \int { dz\over z }{ d \bar z \over { \bar z } } 
                        O^{I_n} (z) \bar O^{I_n} ( \bar z ) |0>  \cr  }}
We can define the z-independent 
operator
\eqn\alphi{ \alpha_{-n} =  
                      \sum_{I_n}\int { dz\over z } 
                       O^{I_n} (z) \int { d \bar z \over { \bar z } }
                                             \bar O^{I_n} ( \bar z ) } 
The dual operator is naturally defined by taking the 
standard CFT dual
\eqn\alphdual{ \alpha_{n} 
= \int {dz \over z} {d \bar z \over { \bar z } } 
       z^{-2 \Delta_n} { \bar z}^{- 2 \bar \Delta_n } 
\alpha_{-n} ( z, \bar z )  } 
This changes the sign of the
oscillator number associated with a conformal 
field. 
Since the expansion of a conformal field
$O^{I_n}$ of weight $\Delta$ looks like 
$ \sum_{m} O^{I_n}_{m}z^{-m-\Delta_n }$, the contour 
integral will extract $O^{I_n}_{-\Delta_n}$. 
Operators  $O^{I_n}_{-\Delta_n-m}$ are descendants 
of chiral primaries for  $m>0$.

\subsec{ Deformed Heisenberg Algebras of the Twist Operators.  }  
 
In the previous subsection, we   discussed deformed 
Heisenberg algebras associated with 
 operators in the untwisted sector of the CFT.  
 These deformed algebras led to the 
nilpotence of the generators of the chiral ring. 
They may also be expected  to  lead 
to a finiteness in  the number of generators of the chiral 
 ring. 
 We will see that the same qualitative properties can be  deduced from 
 the deformed Heisenberg algebras associated with oscillators 
coming from the twisted sector. We will use here 
some of these facts about twist operators derived
 in the previous subsection. 

We can write the relations of the Heisenberg 
algebra, taking $n$ even :  
\eqn\heis{\eqalign{ 
&   [ \alpha_{n} , \alpha_{-n} ] \cr 
&   =  \sum_{I_n, I_n^{\prime}  } 
 [   \bar O^{\dagger I_n} O^{ \dagger I_n } , O^{I^{\prime}_n} \bar
  O^{I_n^{\prime}} ] 
\cr
&  = \{ O^{\dagger I_n } , O^{I^{\prime}_n} \} 
\{ \bar O^{\dagger I_n }, \bar O^{I^{\prime}_n}  \} 
     -   \{  O^{\dagger I_n }  , O^{I^{\prime}_n}    \}  
      \bar O^{I^{\prime}_n}  \bar O^{\dagger I_n }
     -    \{    \bar O^{\dagger I_n }  , \bar O^{I^{\prime}_n}  \} 
      O^{I^{\prime}_n} O^{\dagger I_n } \cr 
&= \delta_{I_n I_n^{\prime} }  \delta_{I_n I_n^{\prime} } 
    - \delta_{I_n I_n^{\prime} } \bar O^{I^{\prime}_n} 
     \bar O^{\dagger I_n }
 -  \delta_{I_n I_n^{\prime} } 
            O^{I^{\prime}_n}  O^{\dagger I_n }  \cr
& = C_{n,N} - ( \bar O^{I_n}  \bar O^{\dagger I_n } +
 O^{I_n} O^{\dagger I_n } ) .
\cr  }} 
$C_{n,N}$ is the (positive)  number 
of elements  in the conjugacy class in $S_N$
characterized by one non-trivial cycle of length 
$n$. To leading order in large $N$
this coefficient behaves like $N^n$. 
The separate purely chiral and purely 
anti-chiral commutators are computed 
using contour integrals.
\eqn\contcomp{\eqalign{
 \{ O^{I_n} , O^{\dagger I^{\prime}_n} \} &= - \int dw w^{ 2 \Delta_n - 1 }
 \int dz ( 1 + { (z-w)\over w} )^{-1}   O^{I_n}(z)O^{*I^{\prime}_n} (w)  \cr
 &=  \delta_{I_n I^{\prime}_n} \cr } }
 After redefining the normalization of these 
oscillators by dividing by this order of a conjugacy 
class we can get $1$ as the leading term. 
The result for the deformed Heisenberg 
algebra is 
\eqn\heisres{\eqalign{ 
  [ \alpha_{n} , \alpha_{-n} ] 
    = 1  - { 1 \over C_{n,N} } (    \bar O^{I_n} \bar {O^{\dagger}}^{I_n}+ 
                        O^{I_n} {O^{I_n}}^{\dagger}  ) \cr }} 
The operator  $\bar O^{I_n} \bar {O^{\dagger}}^{I_n} + O^{I_n}
 {O^{I_n}}^{\dagger}  $ 
 is positive and appears with a minus sign in front of it
( for operators  $\alpha_{-n}$ with $n$ odd the sign is different 
 and cutoffs could be derived by arguments using more detailed
 properties of these operators). 
 It can be expressed as a sum, with positive coefficients, 
 of  number operators 
 for the various oscillators in the theory. 
This can be done along the lines 
of the calculation in section 2. 
 The positivity of the coefficients in front 
of the number operators follows from the fact that 
they are related to the numbers which appear 
in the class multiplication algebra of the symmetric group. 
We pause to explain what we mean by this. 
Let $T_a$ be a conjugacy class in the symmetric group
$S_N$. Define in the group algebra of $S_N$ the 
sum of all the elements in $T_a$, and denote it, by a slight 
abuse of notation, $T_a$. For two conjugacy classes
$T_a$ and $T_b$ we have a product which can be expanded 
\eqn\exp{ T_a T_b = C_{ab}^{c} T_c }    
The coefficients $C_{ab}^c $ are positive. 
 Thus there are various constraints on the
 the values of the number  operators coming from 
 the requirement that $< a a^{\dagger} >$ is positive.  
 This puts restrictions on the values 
 that the number operators can take on the 
 states obtained by  acting with the chiral primary 
 operators on the vacuum.

It is very interesting that the 
commutators between oscillators corresponding 
to permutations of different cycle lengths are 
also non-trivial, although they vanish at large $N$.   
\eqn\mixheis{ 
   [ \alpha_{ l_1 } , \alpha_{-l_2} ] 
    =  a_1  \alpha_{- ( l_2 - l_1 + 1) }   +
       a_2 \alpha_{ - ( k_1, k_2 \cdots ) }  + ... } 
We have on the right hand side 
oscillators corresponding to permutations 
for which the sum of cycle lengths adds up to 
$l_2 - l_1$.

Relations of the type \mixheis\ 
imply that underlying this system 
there should be a description given by a Hamiltonian
of the 
form 
\eqn\Ham{ H = H_0 + \sum_{ijk} C_{ijk} \alpha_i \alpha_j \alpha_k + \cdots } 
The $H_0$ is a sum related to the 
deformed Heisenberg algebras 
 associated to each individual CFT 
chiral primary operator ( and descendants ).  
The coefficients $C_{ijk}$ are  be 
determined by group theory together with 
the intersection form on the manifold $X$. 
 More precisely we recall that the index $i$
 is really a double index $(i_1 i_2)$, 
 $i_1$  
 labeling  cycle lengths and $i_2$ labeling 
 forms on the manifold $X$. The coefficients 
 $C_{ijk}$ may be expected to decompose
 into a product  $C_{i_1j_1k_1}$ 
 which are struture constants of symmetric group
 multiplication,   and $C_{i_2j_2k_2}$ which 
 are intersection numbers coming from the 
 manifold $X$. Interesting questions 
 relate to the inclusion of non-chiral primaries 
 in \Ham.

\subsec{   Generators of the chiral ring.   }  

The single particle states in gravity 
 are identified with the generators of the 
 chiral ring of the CFT. Multiparticle 
 states are associated with chiral primary operators 
 which are products of the generators \deboer. 
It  will be important to clarify the 
nature of the cutoffs on generators of the 
chiral ring which comes from the 
orbifold CFT. For concreteness we will 
discuss these relations in the context 
of the CFT of $S^N(T^4)$. For this purpose 
we will write down some explicit formulae 
for the chiral primary operators
 associated with the twisted sectors of the $S_N$ 
orbifold theory. 
 We argue that the 
 simple  pattern that emerges from such a discussion 
 extends to the CFT for $S^N(K3)$.

 Among the generators of the chiral primaries 
 are $\alpha_{-n}^{0,0}$. The operator 
 $\alpha_{-1}^{0,0}$ operator corresponds to 
 the vacuum state in CFT, so it does not correspond to 
 a particle excitation in gravity. 
 In an $S_N$ orbifold CFT the index $n$ is clearly 
 cutoff at $N$. The $SU(2) \times SU(2)$ quantum numbers
$(2J_L, 2J_R)$ 
 of $ \alpha_{-N}^{(0,0)}$ are $(N-1,N-1)$. 

 The CFT also contains operators $\alpha_{-n}^{p,q}$ 
 for any $(p,q)$ and any $n$ extending up to $N$. 
 However some of these  have the interesting property 
 that they can be written in terms of products of generators. 
 It will turn out that these relations have a remarkably 
 simple consequence. The cutoffs can be described
 by saying that, for the generators, we have a bound 
\eqn\precut{  
 \hbox { max } ( 2J_L, 2J_R ) = (N-1). }  
 Note that this cutoff is stronger 
 than the one based on the unitarity of
 the superconformal algebra \lvw\ which
 says that $\hbox { max } ( 2J_L, 2J_R ) = c/3 = 2N$. 
 In the next section we will explain that 
 the precise cutoff \precut\ 
 follows from a simple spacetime argument 
 based on the picture that quantum effects 
 result in a q-deformation of spacetime, 
 with $q = e^{i \pi \over {N+1}}$. 
 
 The first set of relevant relations takes the form 
\eqn\reli{ \alpha_{-N}^{(p,q)}  = \alpha_{-1}^{(p,q) } 
\alpha_{-N}^{(0,0) }.  } 
 For $(p=1,q=0)$ and $(p=0, q=1)$, this follows trivially from the 
 definitions : 
\eqn\reliei{\eqalign{ 
  \alpha_{-1}^{(1,0)} &= \sum_{i=1}^{N} \psi^{a}_{i} \cr    
 \alpha_{-N}^{(1,0)}  &=  \alpha_{-N}(X^{i}, \psi^{i} )
        \sum_{i=1}^{N}   \psi^{a}_{i} \cr 
                      & =   \alpha_{-N}^{(0,0)} \alpha_{-1}^{(1,0)}   \cr }}
The next case will illustrate the form of the argument 
 which can be used in all the cases in \reli.  
\eqn\relieii{\eqalign{
 \alpha_{-N}^{(1,1)} &= 
  ( \sum_{i=1}^{N} \psi_{i}^{a} \bar \psi_{i}^{ \bar b }) 
    \alpha_{-N}^{(0,0)} \cr 
 & = ( \sum_{k=0}^{N-1} \psi_{(k)}^{a} \bar \psi_{(N-k)}^{ \bar b} ) 
           \alpha_{-N}^{(0,0)}, \cr }}
where the objects with simple $Z_N$ transformation 
 properties $\psi_{(k)}$ have been defined
in \lincomb.  
Now we have to recall that the product 
 that defines the relations between the 
 generators of the chiral ring involves 
taking the OPE of the CFT fields and   
 analysing the leading $z$-independent term. 
 The $k=0$ term certainly gives a contribution
 to the leading constant term. 
However the other terms give terms that 
 vanish in the $z=0$ limit. 
But the $k=0$ term is immediately identified  
 with the first line of \relieii. 
The case of $K3$ involves a slightly more complicated
 form for these operators. We will have, in general,
some function of $X$ appearing in the form 
$f(X)\psi \bar \psi$. The similar decomposition 
into $Z_N$ invariants will have a sum 
$\sum_{k_1,k_2,k_3} f_{(k_1)} \psi_{(k_1)} \psi_{(k_3)}$, 
where $k_1 + k_2 + k_3 = 0 (mod \quad  N)$. 
For a non-zero leading term in the OPE
 we again have $k_1 = k_2 = k_3 = 0 $.

Now consider the cases either $p$ or $q$ 
 is equal to $2$, but they are not both 
 equal to $2$. 
In these cases we show directly here  that  
$\alpha_{-(N-1)}^{(p,q)} $ can  be written 
in terms of products of other generators. 
The proof in all these cases is essentially
 of the same form, so we will illustrate 
in the case of $\alpha_{-(N-1)}^{(2,1)}$. 
The candidate operators which can give 
$\alpha_{-(N-1)}^{(2,1)}$ are 
\eqn\possib{\eqalign{
&  \alpha_{-(N-1)}^{(2,0)} \alpha_{(-1)}^{(0,1)}  \cr 
&  \alpha_{-(N-1)}^{(1,0)}  \alpha_{(-1)}^{(1,0)}
   \alpha_{(-1)}^{(0,1)}  \cr 
& \alpha_{-(N-1)}^{(0,0)} \alpha_{(-1)}^{(1,0)} 
    \alpha_{(-1)}^{(1,0)} \alpha_{(-1)}^{(0,1)} \cr }}
The set of operators that can appear 
is restricted by conservation of charge 
$(2J_L, 2J_R)$, and by the structure constants 
 of symmetric group multiplication. 
As an illustration of the latter 
constraint, the first possibility is allowed 
because the multiplication of the conjugacy class 
containing one cycle of length $(N-1)$ associated 
with $\alpha_{-(N-1)}^{(2,0)}$ with the identity 
permutation associated with $\alpha_{(-1)}^{(0,1)}$  can give 
the first conjugacy class again which is associated with 
$\alpha_{-(N-1)}^{(2,1)}$.

To simplify notation in this section 
denote the operator 
$ \sigma(  X^{i}_{i_1 \cdots i_n}, \psi^{i}_{ i_1 \cdots i_n  } ) $
 by $\sigma_{i_1 \cdots i_n }$. 
Consider the product 
\eqn\prodgs{ \eqalign{      
   \alpha_{-(N-1)}^{(1,0)}&  \alpha_{-1}^{(1,0)} \alpha_{-1}^{(0,1)} \cr
 & =\sum_{(i_1, \cdots i_{N-1} ) } \snm ( \psi^1_{i_1} + \psi^1_{i_2} 
   + \cdots \psi^{1}_{i_{N-1}} ) \sum_{j} \psi^2_j 
            \sum_{k}  \bar \psi^2_k \cr  
& = \sum_{(i_1 \cdots i_{N-1}) } \snm ( \psi^1_{i_1} + \psi^{1}_{i_2} + \cdots 
          \psi^1_{i_{N-1} } ) ( \psi^2_{i_1} + \cdots \psi^2_{i_{N-1}} ) 
         ( \bar \psi^1_{i_1} + \cdots \bar \psi^1_{i_N-1} ) \cr 
& \qquad + \snm  ( \psi^1_{i_1}  + \cdots  \psi^1_{i_{N-1} } )
         ( \psi^2_{i_1} + \cdots \psi^2_{i_{N-1}} )
         \bar \psi^{1}_{i_N}  \cr 
 & \qquad + \snm ( \psi^1_{i_1}  + \cdots  \psi^1_{i_{N-1} } )
           \psi_{i_N}^{2}     
          ( \bar \psi^1_{i_1}  + \cdots \bar \psi^1_{i_{N-1}}  )      \cr 
  &\qquad + \snm ( \psi^1_{i_1} \cdots  \psi^1_{i_{N-1} } ) \psi^2_{i_N} 
       \bar \psi^1_{i_{N}} \cr 
&= \alpha^{2,1}_{-( N-1)} + 
     ( - \alpha_{- (N-1) }^{(2,1) } + \alpha_{ N-1}^{(2,0)}  
     \alpha_{-1}^{(0,1)}  ) \cr  
&  \qquad (-\alpha_{-(N-1)}^{(2,1)} + \alpha_{-(N-1)}^{ (1,1)} 
       \alpha^{ (1,0) }_{-1} ) \cr 
& \qquad (-\alpha_{-(N-1)}^{(2,1)} +
 \alpha_{-(N-1)}^{ (1,0)} \alpha^{(1,1) }_{-1})
\cr}}
So all the terms which are allowed by $S_N$ symmetry 
together with the $U(1) $ charge conservation 
 appear in the above equation. 

This can be rewritten as 
\eqn\expto{\eqalign{ 
2 \alpha_{-(N-1)}^{(2,1)} &= 
  \alpha_{-(N-1)}^{(1,0)} \alpha_{-1}^{(1,0)} \alpha_{-1}^{(0,1)} 
  +  \alpha_{-(N-1)}^{(2,0)} \alpha_{-1}^{(0,1)}\cr
& \qquad  +  \alpha_{-(N-1)}^{(1,1)} \alpha_{-1}^{(1,0)}
  + \alpha_{-(N-1)}^{(1,0)} \alpha_{-1}^{(1,1)} \cr }}

Finally we outline the proof of an equation  
expressing the operator $\alpha_{-(N-1)}^{(2,2)}$
in terms of other operators. 
We can start with a product of 
the form $\alpha_{-N}^{(0,0)} \alpha_{-2}^{(0,0)}$. 
Charge conservation and the structure of the symmetric group
  allows products of the form 
\eqn\cands{\eqalign{ 
 & \alpha_{-(N-1)}^{(2,2)} \cr  
   & \alpha_{-(N-k)}^{(p_1,q_1)} \alpha_{-k}^{(p_2, q_2)}  \cr}}
where $k$ is an odd number. 
 A term like $\alpha_{-(N)}^{(1,1)}$ is allowed 
 by left and right $U(1)$ charge conservation, 
 but it is not allowed by the multiplication 
 law of the symmetric group, since the product 
 of a permutation with a single non-trivial 
 cycle of length  $(N-1) $ and another with length  $2$
is even or odd depending on the sign of 
 $(-1)^N$ whereas the  parity of the permutation 
 with cycle of length $N$ is $(-1)^{(N-1)}$. 
 Thus we may expect a relation with the product 
    $\alpha_{-N}^{0,0} \alpha_{-2}^{0,0}$ on the left 
 and on the right hand side a term proportional to 
$ \alpha_{-(N-1)}^{(2,2)}$ and other terms 
 proportional to products of $\alpha$. No other 
 term with a single $\alpha$ can appear. Terms 
 corresponding to smaller cycles would  require 
 associated forms of degrees which exceed $2$. 
As a result, $\alpha_{-(N-1)}^{(2,2)}$ can be written 
 in terms of products of other $\alpha$'s, 
 so the highest $n$ allowed for $\alpha_{-n}^{(2,2)}$
 among the generators of the chiral ring 
 has $ n = N-2$ and charges $2J_L= 2J_R = N-1$.

\newsec{ Spacetime interpretation of the Exclusion principle } 

We now proceed to  discuss the 
spacetime interpretation of the exclusion principle.
The first  main result in this section is that the 
detailed form of the cutoffs on the 
generators of the chiral ring has a simple
interpretation in terms of our proposal 
that spacetime is deformed from 
$ADS_3 \times S^3 $ to $ SU_q(1,1)\times SU_q(2)$. 
In section 3.1,  we review some properties 
of $SU_q(2)$ which follow from the 
consideration of study of unitary representations
of $U_qSU(2)$. We then identify the $q$ parameter 
associated with the deformed geometry, by comparing to the 
discussion in the last part of section 2. 
In section 3.2, we present arguments in favour of the 
deformation of the non-compact part of spacetime 
( in addition to the deformation of the compact 
 part which played  a direct role in the discussion of the chiral 
 primaries). In section 3.3, we show that the 
 deformation to $SU_q(1,1)$ plays an important role 
 in the discussion of the cutoffs on descendants of chiral primaries. 
 Section 3.4 discusses holography in the
 $q$-deformed context. 
In Section 3.5 we observe the equality 
of the geometrical  $q$-parameter with the one obtained from deformed 
Heisenberg algebras in section 2, and we discuss its implications.
In  section 3.6  we outline some steps towards 
 an explicit derivation of a non-commutative 
space-time coordinates from the orbifold CFT. 
In the final part we begin a very preliminary 
 discussion of other ADS backgrounds.

\subsec{ Non-commutative $S^3$ and the cutoff on chiral primaries. }  

The  above cutoffs
 on the left and and right $su(2)$
 quantum numbers of the generators of the chiral 
ring have  an interpretation in terms of KK reduction on 
 a non-commutative deformation of  $S^3= SU(2)$ to 
$SU_q(2)$ with $q = e^{i\pi \over { N+1}}$. 
We have the oscillators $\alpha_{-l}^{(p,q)}$ 
associated 
with single particle states of supergravity on $ADS_3  \times S^3$ 
carrying $SU(2)$ spins $ 2J_L = l-1+p$ and  $2J_R = l-1 + q$.
 The KK reduction on $S^3$ gives
states characterized by the reps of $SU(2) = S^3$ because functions 
on $SU(2)$ can be expanded in terms of matrix elements of representations 
of the group. The manifold $S^3$ being isomorphic 
 to the group $SU(2)$  admits a 
 natural deformation to $SU(2)_q$ \worcomp\ which preserves 
 many  group  structures. Such a manifold admits 
 $SU_q(2)\times SU(2)_q$ symmetry.

The universal enveloping  algebra 
$U_qSL(2,C)$ is generated by $H,X_+, X_-$ 
with the relations 
\eqn\liealg{\eqalign{&   [ H, X_{\pm } ] = \pm X_{\pm } \cr  
                     &   [ X_+, X_- ] = 
                  { ( q^{2H} - q^{-2H} ) \over { q - q^{-1} } } \cr   }}
The algebra is also equipped with a co-product
 $ \Delta : A \rightarrow A \otimes A $, where $A = U_q SL(2,C)$. 
 As an example 
\eqn\cop{ \Delta (X_+ ) = X_{+} \otimes q^H + q^{-H} \otimes X_{+} } 
Note that the coproduct is not invariant 
 under permutation of the first and second copies 
 of the algebra. The two co-products are related by the R-matrix. 
This non-cocommutativity leads to the fact that 
 the manifolds constructed from these algebras 
 have  non-commutative spacetime coordinates. 

Specifying a real form of the Lie algebra, 
e.g. a compact form $SU(2)_q$ or a non-compact form $SU(1,1)_q$ 
involves the choice of an involution $*$ which 
is a map from the algebra to itself, satisfying 
\eqn\inv{ * ( ab) = (*a) (*b) } 
The q-algebra is also equipped with 
a coproduct $ \Delta : A \rightarrow A \otimes A $.
One also requires a compatibility 
between the conjugation and the 
coproduct which can take one of two forms. 
\eqn\invcomp{\eqalign{&   (I) ~ * \Delta = \Delta *  \cr 
                      &   (II) ~  * \Delta =   \Delta^{\prime }  * }}
If we require I the $SU(2)$ involution 
exists for real $q$ but not $ |q| = 1$. 
If we require II, the compact involution 
exists for roots of unity.  We will discuss involutions 
which go to those related to $SU(1,1)$ in the limit 
$q \rightarrow 1 $ in section 4.

Given a definition 
of conjugation, compatible with the q-deformed
algebra we can discuss unitarity of representations. 
The set of unitary highest weight reps of $SU(2)_q$ 
for $q = e^{ i \pi\over {k+2}}$ is truncated to those 
having   $2j  \le k $. 
The algebra of functions 
on the non-commutative manifold  $SU(2)_q$ at roots 
of unity is given by the matrix elements of a finite set of representations. 

The existence of $SU(2)_q \times SU(2)_q$ 
symmetry associated with the $q$-sphere 
 implies that KK reduction on the non-commutative 
sphere should correspond to a representations
which have $(2J_L, 2J_R)$ which are both 
 bounded by $2k$. Equivalently $ max( 2J_L, 2J_R ) \le k$. 
This is precisely the form of the cutoffs 
 we get from the above discussion, with 
a value of $q= e^{i\pi \over {N + 1 }}$. 

In this way we are lead to a suggestion that some of the 
finite $N $ constraints in the $ADS_3 \times S^3$
background can be captured by Kaluza-Klein reduction on a  
non-commutative 3-sphere $SU(2)_q$.

\subsec{ Non-commutative $ADS_3$ and $SU_q(1,1)$ }

We have argued that finite $N$ effects 
lead to a deformation of the 
classical $S^3$ of spacetime to 
a $q$-deformed $S^3$. 
We will  argue here  that 
the finite $N$ effects also deform 
the $ADS_3$  part of spacetime into a non-commutative 
manifold. The definition of the non-compact quantum group 
has a number of subtleties. We  use the 
proposed relation between the non-compact quantum group 
and the CFT  to learn something 
about the correct properties of the non-compact quantum group.

 In the classical case, KK reduction 
 of scalars on the sphere $S^3$ 
 leads to states which correspond to the chiral primaries 
 which are finite dimensional highest weight 
 reps of $SU(2)$ and infinite-dimensional 
 highest weight reps of $SU(1,1)$. The highest 
 weight $h$ of the $SU(1,1)$ is related 
 to the highest weight $j$ of the $SU(2)$ 
 by $h = j/2 $. This follows,  from the 
spacetime point of view, by  writing down 
 a sum of Laplacians on the sphere 
 and the ADS space, and taking into account 
 the mixing between the scalars and some modes 
 of the antisymmetric tensor fields. 
The net effect is to obtain a relation 
 of the form 
\eqn\relcas{ 4h(h-1) = j ( j-2  ) } 
 which allows a solution $h = j/2 $. 
 We expect that the 
 KK reduction of the $S^3_q$  involves 
 the q-Casimir rather than the ordinary 
 Casimir. A condition of the 
 form $h = j/2$ present in CFT,  will only follow,  if the 
 $SU(1,1)$ is also q-deformed.

This gives an argument that the $AdS3$ ( $= SU(1,1)$ ) space
is 
also q-deformed. 
Deformations of the non-compact real forms 
are significantly more subtle than the compact form. 
For example, one  non-compact form of 
\matsue\ exists
for real $q \ne 1$ at the level 
of Hopf algebra but does not have the right properties
to lead to a quantum group as a non-commutative manifold  \wor. 
This subtlety, fortunately 
does not concern us 
 since we have a q-deformed $Sl(2)$ for 
$q$ equaling a root of unity.

Deforming the algebra of bounded functions 
 on the non-compact group will involve 
replacing a set of classical unitary representations 
of $SU(1,1)$ with some class of unitary representations 
of  $SU(1,1)_q$. 
Typically in quantum group-CFT relations 
the structure of the quantum group representations 
is similar to those appearing in the  CFT. 
In this  case the conformal 
 field theory contains $SU(1,1)$ representations 
 which have a highest weight. 
So we need a definition of the 
non-compact form which admits the discrete 
series representations. 

Representations of $U_q SU(1,1)$
at roots of unity have been studied in  \suz\ and \har.   
There are analogs of the unitary discrete series.
This is an encouraging sign that their reps. 
might be useful in a definition of the 
q-deformed $SU(1,1)$ manifold relevant here, 
since the proposed CFT dual has discrete series 
representations of $SU(1,1)$.    

The definition of the q-deformed $ADS_3$ used above and in  
sections  3.3,  3.4  is based on $U_q SU(1,1)$, where the *
operation satisfies the rel. (II) of \invcomp\ between 
 the conjugation and the  co-product.
 A different  possibility which we have not explored 
 much is to use $U_qSl(2,R)$ \matsue\sch\ since 
 the classical isomorphism between $SU(1,1)$ and $SL(2,R)$
 does not extend to the case $q \ne 1$. 
 While the discussion above and in   sections 3.3, 3.4 finds
  evidence in favour of $U_qSU(1,1)$, it remains 
 an interesting question to find analogous 
 applications of $U_q Sl(2,R)$.

\subsec{ Discussion of descendants } 
While we focused on chiral primary operators
( and their $SU(2) \times SU(1,1)$ descendants )
above,  it is interesting to consider  
 other operators which are obtained from these
 by acting with the superalgebra. 
 We can use the notation of \deger\  to describe 
 the set of reps. 

 There are reps. $D^{(l_1,l_2)}_{(E_0,s)} (n,k)$
 The left and right $SU(2)$ quantum numbers 
 are $2J_L = (l_1 + l_2 )$
and  $2J_R=( l_1 - l_2 ) $. The left 
 and right $SU(1,1)$ quantum numbers 
 are given by $ 2h = (E_0 + s)$ and 
$2 \bar h = ( E_0 -s )$. The numbers
 $(n,k)$ are quantum numbers under 
 a symmetry group  
 $SO(4) \times SO(n_T)$, 
 where $n_T = 21 $ for $K_3$, and 
$n_T = 5 $ for $T^4$. 

The oscillators $\alpha_{-n}^{p,q}$ 
 have quantum numbers $ (2J_L = n-1 + p, 2J_R = n -1 + q )$, 
 and $2h_L = p + q + n-1 $ and $2h_R = p - q + n-1 $. 
 This allows an identification 
 of these $\alpha$ oscillators with 
 Figs. 1, 2, and 3 of \deger. 
 The oscillator $ \alpha_{- ( l+2 ) }^ {(2,0) }$
 corresponds to $D^{(l+2,1)}_{( l+2, 1 ) }$ 
of fig. 1. 
 The oscillator $ \alpha^{2,2}_{-(l+1)} $ corresponds to 
 $D^{(l+2,0)}_{(l+2,0)}$ of fig. 2. 
The oscillators $\alpha_{-(l+2)}^{(1,1)}$
 and $\alpha_{-(l+3)}^{(0,0)}$ are associated 
 with $D^{(l+2,0)}_{(l+2,0)}$ of fig. 3.

The discussion in section 3 
 implies that there is a cutoff in the
set of D's in Fig. 1, given by  $ \hbox {max}  ( l+2 ) = (N-2) $.
For Fig. 2 we have $ \hbox {max} ( l+1 ) =( N-2)$.   
For Fig. 3, we have $  \hbox {max}  (l+2) =( N-1)$. 

Once we have decided how the 
chiral primaries are cutoff, SUSY requires
 the associated descendants to be cutoff correspondingly. 
It is interesting to ask nevertheless 
how far we can predict the cutoffs on the descendants 
 using only 
 the quantum group symmetry $SU_q(1,1) \times SU_q (2)$. 
For concreteness we discuss Fig. 2. 
To carry out the argument we will need
 the fact both $SU(2)$ and $SU(1,1)$ are 
q-deformed. For $SU(2)$ we have the usual 
$2J \le (N - 1) $, where $2J$ denotes 
the larger of $2J_L $ and $2J_R$. 
For $SU_q(1,1)$ we have the cutoff 
 $2h \le N+1$, using the result of 
 \har\ on the  consistent restriction to a 
set of unitary $SU_q (1,1)$ reps. 

Using these cutoffs we can correctly predict 
the cutoffs on the entire diamond of \deger\ 
except for the one rep. in the middle. 
The correct way to cut-off the set of reps 
 is of course the one compatible with 
 SUSY, and the extra state in the middle, 
 allowed by unitarity of $SU_q(1,1)$ 
 and $SU_q(2)$ should not appear. The consistent 
 truncation should follow from constraints based 
 on larger algebra. The simplest candidate 
 is $SU_q(2|1,1)$, but some larger algebra containing
 this as a subalgebra could be involved. 
 It is very appealing nevertheless that the  idea
 of q-deformation of the bosonic geometry alone
 is sufficiently powerful to predict the 
 correct cutoffs for almost the entire figure.

\subsec{Boundary of q-SU(1,1) and holography. }

We are arguing here that 
the spacetime gravity theory which 
can be reconstructed from the CFT, 
has finite $N$ effects which can be understood in terms 
of a non-commutative spacetime. 
Among other things, we have argued that 
a  non-commutative $q$-deformation of the $SU(1,1)$ manifold is relevant.
An important feature of the ADS-CFT correspondence,
emphasized in \withol\ is that it involves 
a theory at the boundary being dual to the bulk gravity theory.
In all the ADS/CFT examples of \malda\ the boundary 
 theory lives on a space $X$ which appears as a factor 
 in a boundary   space of the form $X \times Y$. Similarly in 
 \gopvaf\ we have string theory on a manifold
  which at infinity takes the form $ S^3 \times S^2$ 
  and is dual to a gauge theory on a manifold $S^3$.
  More generally we may expect $X$ to appear as the base
  of a fibration at infinity. 
   Another very simple string-gauge theory correspondence 
    fits this general paradigm. The string theory 
     of 2D QCD  on a Riemann surface $\Sigma$ 
     \grta\cmr\hortwod\ is related to a topological string theory 
      on  the cotangent bundle $T^* \Sigma$ \dm. 
      The boundary of the  $R^2$ fibre is $S^1$ so the 
       boundary of the target is an $S^1$ fibration over $\Sigma$. 
      
 In this non-commutative context, 
 an interesting property of the $q-SU(1,1)$ 
 manifold   follows from the assumption  that 
 the standard holographic picture continues to hold.
 Assuming that holography continues to hold, 
 the $q-SU(1,1)$ manifold should have a boundary 
 containing an ordinary  commutative
 torus, on which the dual finite N orbifold  CFT is defined. 
   Indeed, one has \suz\
 a tensor product structure in the space of unitary reps
 of $U_q(SU(1,1))$ of the form 
\eqn\produniv{ 
 U_q(SU(1,1))=U_q(SU(2)) \times U ( SU(1,1) ) 
} 
This means that if we define the non-commutative  deformed 
algebra
of functions on q-$SU(1,1)$, using this class of 
unitary representations, we have a product structure on the
q-deformed manifold:
$$
  SU_q(1,1)=SU(1,1) \times SU_q(2)
$$
allowing us to find the boundary of q-$SU(1,1)$. It is equal to the
boundary of $SU(1,1)$ which  is an ordinary torus times a 
compact q-manifold. ( The factorization of the quantum 
 geometry has been discussed in the context of Liouville
 theory in \suzi ).   Consequently the
q-deformation of the AdS space 
agrees with our expectations based on holography.
The fact that the ordinary Riemann 
surface appears at the boundary allows us to 
hope that the precise expression  
of holography of the form expressed in \wit\ as
\eqn\holeq{ \left\langle  exp \int \phi_0 {\cal O } \right\rangle_{CFT}  
                  = Z_{S} ( \phi_0 ) }  
will have a q-generalization where 
$Z_{S}$ is obtained from a theory on the  non-commutative 
space $ SU_q(1,1)$, and the left hand side is 
computed from orbifold CFT on an ordinary 
Riemann surface.

The above discussion may raise a small 
puzzle which has a simple solution in this physical 
context. The derivation of the product structure 
\produniv\ involves a somewhat unintuitive 
procedure of dropping the operators $X_{\pm}^{(N+1)}  $
and adding back the operators  
 ${ X_{\pm}^{(N+1)} \over { [N+1]!}}$ \suz.
In such a discussion the full set of reps 
$V_{d,z}$ ( in the notation of \har) 
for arbitrary $z$  plays an important role. 
For the discussion of particle excitations around the ADS 
background, a procedure which 
  restricts attention ( as  for example in \har\ )
 to the highest weights associated to $z=1$  is more appropriate. 
As discussed in section 
2, the $X_+$ operators are closely related 
 to particle creation operators, and it makes 
sense to include the ${ X_{\pm}^{(N+1) } \over { [N+1]!}}$
if we allow ourselves to change the background, 
but not if we are discussing 
the physics around a fixed background as in the previous section. 
The freedom to add the extra operators should also be of help 
in setting up a spacetime description which 
allow accounting for the black hole entropy, a question 
which received an interesting discussion recently in \dbii.  
For the purposes of discussing the global structure of
spacetime, as we are doing here, 
 it is natural to include the extra operators. 
Equivalently, the orbifold CFT contains 
 the states relevant for black holes, 
so it should appear as the boundary 
 of a quantum spacetime whose definition 
 allows for operators which can create very heavy objects.

\subsec{ Nilpotence, finiteness and the q-deformation parameter  } 

 For quantum group at a root of unity, 
 the representation ring truncates to a finite set. 
 For $q = e^{ i \pi \over {k+2}}$ we get 
 a restriction on the $SU(2)$  spins   $2j \le k $. 
 We saw that the cut-offs on the spectrum of 
 generators $ \alpha_{-l}^{(p,q)}$  of the chiral ring 
 takes a simple form $2j \le ( N - 1 ) $.  
 The generators of the chiral ring are associated 
 with single particle states in space-time
 and their cut-off is directly related to the 
 $q$-deformation parameter associated with 
 the geometrical deformation of $SU(2) \times SU(1,1)$ 
 manifold to the non-commutative manifold 
 $SU_q (2) \times SU_q (1,1)$. 
  Earlier in section 2, we have argued for 
 a $q$-deformation of the field oscillators
and we derived in this context a deformed 
algebra $Sl_q(2)$. This deformation was directly 
 related to the nilpotence of the generators of the chiral 
 ring.  
 In both cases we
 find the same value of $q = e^{i \pi \over { N +1}}$. 
While the truncation of the oscillator 
index can be interpreted simply in terms of the 
$q$-deformation of the geometry of background spacetime, the
deformed Heisenberg algebras appear more directly
 related to the deformation of the dynamics 
 of fluctuations around the background. 
In fact a number of physicists have explored
the possibility of doing field theory in the non-commutative 
context by deforming the oscillator algebras of ordinary 
field theory, see for example \fink. 
The equality of the $q$-parameters obtained 
 from these two different routes suggests that there are tight 
 consistency conditions relating
 the deformation of space-time geometry 
 and the deformation of field oscillator
 algebras.

 An important 
 role is played in non-commutative geometry of $SU_q(2)$
 as well as in deformed oscillator algebras by Hopf algebras. 
 This leads to the suggestion that it might be possible to associate 
 a   Hopf algebra structure  to the entire chiral ring 
 and that such a structure 
 will have an interesting role in characterizing 
 the consistency conditions governing the deformed 
 dynamics of fields on a deformed spacetime.

\subsec{Towards 
 a construction of non-commutative 
spacetime from  the CFT. } 

Non-commutative $SU(2)$ manifold  appears naturally 
 in connection with $SU(2)$ WZW CFT, which has been 
interpreted in terms of particle moving 
on $SU(2)_q$ \babel\ldf.  An intuitive understanding of the 
q-deformation of $S^3$ in terms of an effective 
 spacetime coordinate can be given. This would be 
 related to the 
  four dimensional coordinate of \gubs, given as 
\eqn\matx{ x^m = \bar \Psi \gamma^{m} \Psi } 
 which appeared in the calculations of the absorption
cross-sections using the effective string model.

  Consider the 
 $SU(2)_L \times SU(2)_R)$ 
CFT symmetry currents built in terms of fermions as
$$
\eqalign{ J^A (z) & = \Psi^+ \cdot{t}^A \cdot \Psi (z)\cr
\bar{J}^A (\bar{z} ) & = \bar{\Psi}^+ \cdot t^A \cdot \bar{\Psi} (\bar{z} ) }
$$
Here we have the left and right moving Dirac components respectively.
These can
be written as linear combinations of the basic Majorana fermions
$$
\eqalign {\psi^{a\alpha}_{i} (z) \qquad  
\qquad \bar{\psi}_{i}^{\dot{a}\dot{\alpha}}(\bar z )
}
$$
As is now standard $a=1,2$ represent world sheet indices  of the conformal 
field theory fermions and $\alpha , \dot{\alpha}  = 1,2 $, the 
$SU(2) \times SU(2)$ indices.
We now consider a collective coordinate for the KK space.  
An effective coordinate of $S^3$ can be identified from the 
group element in a WZW type construction.  In general
$$
g (z,\bar{z} ) = g_L (z) g_R (\bar{z} )
$$
We now use the well known fact that a global mode exists in the WZW theory.  It is defined through the monodromy
$$
V_q = g^{-1} (z) g (ze^{2\pi i } )
$$
In terms of the current
$$
V_q = P e^{\oint J}
$$
It follows from the fact that the current $J$ obeys 
an $SU(2)$ Kac-Moody algebra  that the global $U_q$ 
degree of freedom obeys a $q$-deformed set of commutators. 
The appearance of a q-group structure is a typical phenomena in 
these theories and was studied extensively in the literature\babel\ldf.
For  SU(2)  the value of the deformation is given by 
$q = e^{ i \pi\over {k+2}}$ 
where $k$ represents the level of the current algebra. 
In the present
case the superconformal chiral algebra contains a
an $SU(2)$ current algebra with   $k = N $ and 
$$
q = e^{{ i \pi\over  N + 2}}
$$
is the related q-deformation parameter. 

The $SU(2)_q$ coordinate $U_q$ is a conjugate to the monodromy $V_q$.
  In the discussion of $SU(2)$ WZW 
  one actually identifies the left and right monodromies.
  This is also appropriate for
 the present interpretation and identification 
 of a center of mass coordinate on $S_3$.  
  The requirement of the single-valuedness
of the correlation functions 
of the chiral primaries leads to a 
correlation between the Monodromy Matrices 
on the left and the right. 
The coordinates $U_q$ belong to the group $SU(2)_q$ 
and the dynamics on this space is jut that of a $q$-deformed top.

 
 To summarize  starting with the 
 $SU(2)$ level $ k = N $  algebra 
 which appears as
 a subalgebra of the superconformal algebra 
 an explicit matrix obeying the commutation relations 
 of $SU(2)_q$, with $q = e^{ i \pi \over { N + 2  }}$
 provides a candidate 
 for a non-commutative $S^3$ as 
 part of the base space for the spacetime. 
 This is not the same 
 parameter that underlies 
 the truncation of the generators of the chiral ring,  
 or the deformed Heisenberg algebra we focused 
 on in section 2. The  operators of interest, 
 the chiral primaries, have simple
 transformation properties under the $SU(2)$ global 
 symmetry of the CFT but do not necessarily have simple 
 properties under the $SU(2)$ current algebra ( e.g they do not 
 have to be primaries of the $SU(2)$ current algebra ). 
This explains why their properties cannot be directly 
 predicted from the current algebra.  
 Nevertheless this line of argument should, with some 
 modification to take into account the quantum effect 
 of shifting $k$ by $1$,  be useful in understanding the 
 space-time coordinate. It is also worth 
 noting that a consistent spacetime picture
 requires, as we argued, a q-deformation 
 of the $SU(1,1)$ part, whereas there is no 
 manifest $SU(1,1)$ current algebra in the 
 orbifold CFT.

\subsec{ The ADS5 case } 

 The conjecture that finite N effects 
 are related to gravity on a non-commutative
 space has interesting consequences in the case 
 of the duality between $ADS_5 \times S^5 $ 
 and $N=4$ super-Yang Mills. 
  As we saw earlier one of the simplest consequences 
 of the non-commutativity is that we have to do 
 KK reduction on a non-commutative sphere as opposed to 
 a commutative one. Quantum groups also 
 suggest a deformation of $S^5$  
 as $SU_q(3)/SU_q(2)$\podles\  
 or perhaps a deformation starting from 
 the quotient $SO(6)/SO(5)$. 
  KK reduction on such a space 
 for $q \sim e^{ i \pi \over N} $
 a root of unity will have the effect 
 that there will be a truncation in the set of 
 representations of $SU(3)$ that appear. 
 In the dual gauge theory this shows up as
 certain dependences between traces of powers 
 of matrices $ \phi^{i}$. We have three complex matrices
 and an $SU(3)$ action. The generating set 
 of independent $SU(N)$ invariant polynomials 
which can be made from these matrices will fall 
 in a certain set of representations
 of $SU(3)$ which will have N-dependent truncations. 
 It will be interesting  to see if these 
 truncations are precisely the ones that appear 
 for $SU(3)$  for a $q$ of the form  $q \sim e^{ i \pi \over N} $. 

\newsec{Summary and Outlook.  } 

We summarize here the main points 
developed above and outline some directions for further research. 

The stringy exclusion principle on the space of chiral primaries 
 implies a number  of consequences for the spectrum of
 quantum  supergravity 
 on $AdS_3 \times S^3$.   From the spacetime point of view 
 there are  qualitatively different 
 aspects of this exclusion principle. 
 The first involves a restriction on the 
  number of particles of one kind. 
 We related this to deformed Heisenberg algebras 
 which we computed from the 
 orbifold CFT. 
 The second involves an upper bound on the number 
 of generators of the chiral ring, which are related to 
single particle states in spacetime. We showed, 
 using an explicit construction of the chiral primaries
 of the orbifold CFT,  
 that this bound takes the simple form of
 a cutoff on the $SU(2)$ spins of the chiral primary. 
 This bound is stronger than the bound we get  
from general arguments based on the unitarity 
 of the superconformal algebra.  
  The precise cutoffs have  a simple interpretation in the 
 context of gravity on a non-commutative deformation of 
 $S^3 \times ADS_3 $ to $ SU_q (2)\times SU_q (1,1)$.
 The hypothesis of a $q$-deformed 
  spacetime with
  $q = e^{i \pi \over { N + 1 } }  $ is, therefore,  more
  accurately  predictive 
 of the precise nature of the cutoffs on the generators  
 of the ring of chiral primary operators 
 coming from the $S_N$ orbifold CFT, 
 than general arguments based on the 
 unitarity of the superconformal algebra,  
 or on the existence of an $SU(2) $ current algebra 
 in the orbifold CFT with $k=N$.

 Further work in  the $ADS_3$ context is needed
 to understand the non-chiral primaries and the 
 super-algebra descendants 
 of the chiral primaries.
 The hypothesis of $q$-deformed 
 spacetime explains a lot about the 
 cutoffs on the superlagebra 
 descendants of chiral primaries, 
 and uses important features about the $q$-deformation 
 of the non-compact part 
 $ADS_3$. 
  We expect 
 that a q-deformation of $SU(1,1|2)$, 
 or perhaps an even larger algebra,  
 will play an interesting role in improving 
 the spacetime understanding of the cutoffs. 
 Given our explicit construction of the chiral primaries, 
 and the remark of \deboer\ that the non-chiral primaries 
 can be obtained from OPE's of the chiral primaries, 
 we expect that the detailed form of the 
  cutoffs on the non-chiral primaries 
 can be deduced directly, by generalizing the 
 arguments used here. 

 The properties of q-deformed $ADS_3$
  also allowed a check that the hypothesis
 of q-deformed spacetime is consistent 
 with holography. This suggests that 
 the relation of \wit\ 
 for correlators in a bulk theory with 
 those of the boundary theory should admit
 a  deformation to the context of a non-commutative 
 spacetime.   
 
Based on the idea that finite $N$ effects 
are also related to a q-deformation of spacetime
we outlined a conjecture on the finite $N$ truncations
 of the spectrum of chiral primary operators
in the case of the correspondence between 
type IIB supergravity  on $ADS_5\times S^5$ and $N=4$ Yang Mills theory. 

 A recent paper of Witten shows that some finite N effects
 can be  understood in terms
 of Chern-Simons actions.  Given the relation between 
 Chern Simons and quantum groups, there should be a 
 close relation between the discussion of \wit\ 
 and the remarks on finite N effects here. A clear elucidation 
 of this would be interesting. It would require a better 
 understanding of the relation between the truncations of the spectrum 
 of chiral primaries studied here and the
 't Hooft fluxes and Wilson loop properties studied in \wit.  
 The investigation of the relation between Wilson loops and local operators 
 of \malfis\ should provide a start.  

 It remains to use the ADS/CFT correspondences 
 to learn more about the theory of fluctuating geometries
 in the non-commutative setting. 
 For example it would be very interesting to write 
 an action for non-commutative gravity on the 
 deformed  $ADS_3 \times S^3$
 space and test its properties using the dual CFT.

 \noindent{\bf Acknowledgements:}
 We are happy to thank for enjoyable and 
 instructive discussions
 Jean Avan,  Rajesh Gopakumar, Yoichi Kazama, Robert de Mello Koch, 
 David Lowe, Mihail Mihailescu,  Jacek Pawelcyk, 
 Phillipe Roche,  Harold Steinacker, Washington  Taylor, Stefan Theisen, 
 Andre van Tonder, Tamiaki Yoneya. We would
 specially like to acknowledge discussions with Julian Lee
 in the early stages of this work. 
This research was supported by DOE grant  DE-FG02/19ER40688-(Task A).

\listrefs 
\end